\documentclass[aps,prl,twocolumn,showpacs,superscriptaddress,floats]{revtex4-1}
\usepackage{graphicx,bm,amsmath,amssymb,natbib,url,epsfig,color,hyperref}

\begin{document}

\title{Detecting Kondo Entanglement by Electron Conductance}
 
\author{Gwangsu Yoo}
\affiliation{Department of Physics, Korea Advanced Institute of Science and Technology, Daejeon 34141, Korea}

\author{S.-S. B. Lee} 
\affiliation{Physics Department, Arnold Sommerfeld Center for Theoretical Physics, and Center for NanoScience, Ludwig-Maximilians-Universit\"{a}t, Theresienstra{\ss}e 37, D-80333 M\"{u}nchen, Germany}

\author{H.-S. Sim}\email[]{hssim@kaist.ac.kr}
\affiliation{Department of Physics, Korea Advanced Institute of Science and Technology, Daejeon 34141, Korea}

\begin{abstract}
Quantum entanglement between an impurity spin and electrons nearby is a key property of the single-channel Kondo effects. We show that the entanglement can be detected by measuring electron conductance through a double quantum dot in an orbital Kondo regime. We derive a relation between the entanglement and the conductance, when the SU(2) spin symmetry of the regime is weakly broken. The relation reflects the universal form of many-body states near the Kondo fixed point.  Using it, the spatial distribution of the entanglement, hence, the Kondo cloud, can be detected, with breaking the symmetry spatially nonuniformly by electrical means.
\end{abstract}

\date{\today}
\maketitle


  

Kondo effects and quantum impurities are central issues of low-dimensional many-body physics~\cite{Kondo, Hewson}. 
In the effects, a local interaction at an impurity leads to macroscopic behavior.
An example is the single-channel Kondo effect, where an impurity spin-$1/2$ is screened by conduction electrons nearby.
The screening electrons spatially extend over a distance (possibly of micrometers), forming a Kondo cloud~\cite{Affleck_cloud1, Barzykin, Sorensen1, Borda, Mitchell,Affleck_attempt1, Affleck_attempt2,Cornaglia,jinhong,Sorensen2,Simon03,Bayat,Eriksson,Holzner09,Hand,Kiselev}.  The screening accompanies entanglement between the impurity and the cloud~\cite{ssblee}. 

For the single-channel Kondo effect, 
a key experimental tool is electron conductance. 
When a quantum dot is coupled with conductors by electron tunneling and hosts an impurity spin with the help of Coulomb repulsion~\cite{Cronenwett, DDG, Pustilnik}, conductance through the dot increases as temperature decreases, reaching the unitary limit~\cite{Wiel} or the Kondo fixed point.
Using this behavior, many universal features of the Kondo effect, such as  scattering phase shift $\pi / 2$~\cite{Nozieres, Hewson_Fermi, Mora,Affleck_CFT1,Gerland00,Takada}, have been identified.

However, the cloud, an essential feature of the Kondo effect, has not been detected, despite efforts~\cite{Affleck_cloud1,Affleck_attempt2,Cornaglia,jinhong,Simon03,Hand}.
The difficulties are associated partially with the fact that the screening accompanies the quantum entanglement.
Detecting entanglement in electron systems is a hard task and has been rarely reported~\cite{Neder},
as it typically requires to see whether mutliparticle correlations are non-classical by using Bell inequalities~\cite{Chtchelkatchev}, multiparticle interferometry~\cite{Samuelsson04,Sim,Neder} or quantum state tomography~\cite{Samuelsson06}.
Its application to the entanglement in the Kondo effect will be even more difficult, since the cloud is a macroscopic object.
Here, we will show that the Kondo entanglement and the cloud can be detected by measuring a single-particle observable of electron conductance.
    

In this work, we consider a quantum dot hosting an impurity spin in the single-channel Kondo regime and analyze the entanglement between the impurity spin and the electron reservoirs of the dot, using the entanglement entropy~\cite{Amico08}.
Using the Fermi-liquid theory~\cite{Nozieres, Hewson_Fermi, Mora} and a bosonization method~\cite{Delft},
we find that the entanglement can be determined from electron conductance through the dot at temperature much lower than the Kondo temperature $T_\textrm{K}$, which is valid even when the SU(2) spin and particle-hole symmetries are weakly broken.
This exemption from measuring multipaticle correlations in determining entanglement comes from the universal form of many-body states near the fixed point.
 
\begin{figure}[b]
\includegraphics[width=\columnwidth]{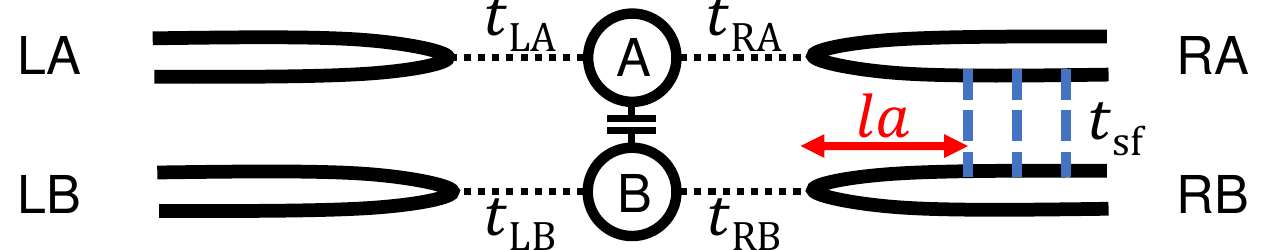}
\caption{A double quantum dot in an orbital Kondo regime. Its degenerate ground states $(n_\textrm{A},n_\textrm{B})=(1,0)$ and $(0,1)$  act as the pseudospin states of the Kondo effect. 
Electron tunneling occurs between dot $\lambda$ $(=\textrm{A}, \textrm{B})$ and its own reservoirs $\eta \lambda$ (dotted lines; $\eta = \textrm{L}, \textrm{R}$), and between reservoirs $\eta \textrm{A}$ and $\eta \textrm{B}$ (dashed) in the region outside distance $l$ from the dot. 
The spatial distribution of the Kondo entanglement is revealed in the $l$ dependence of conductance through the double dot. $l$ can be tuned by electrical gates.  }
\label{setup_AB}
\end{figure}

This finding is useful for detecting the spatial distribution of the entanglement, hence, of Kondo cloud. We propose to use a double quantum dot (see Fig.~\ref{setup_AB}) in an orbital Kondo regime where its orbital degrees of freedom support the pseudospins.  It has the merit that one can break the SU(2) pseudospin symmetry spatially nonuniformly by electrical means. This allows one to detect the spatial distribution of the entanglement by measuring conductance through the double dot. This is confirmed by using the numerical renormalization group method (NRG)~\cite{Weichselbaum07,Bulla08}.
The setup is experimentally feasible as the orbital Kondo effect was observed~\cite{Amasha}.

{\it Entanglement and conductance.---} We first consider a single dot hosting an impurity spin $S=1/2$ in the single-channel Kondo regime, and show that entanglement between the spin $\vec{S}$ and the two (left and right) reservoirs of the dot can be determined from conductance through the dot.
The dot is in Coulomb blockade and has an odd number of electrons. 
The effective Hamiltonian is 
\begin{equation}
\mathcal H = \mathcal H_\textrm{K} - E_Z S_z / \hbar= J\vec{S}\cdot\vec{s} + \mathcal H_\textrm{res} - E_Z S_z/\hbar. \label{Hamiltonian}
\end{equation}
In the Kondo Hamiltonian $ \mathcal H_\textrm{K}$, the impurity spin $\vec{S}$ couples, with strength $J$, to the spin $\vec{s}$ of neighboring reservoir electrons. $\mathcal{H}_\textrm{res}$ describes noninteracting electrons in the reservoirs.
$E_Z S_z / \hbar$ is Zeeman splitting of the dot spin by a magnetic field along $\hat{z}$ axis, and breaks the SU(2) spin symmetry of $\mathcal H_\textrm{K}$; it can also describe other sources (ferromagnetic reservoirs, spin-dependent tunneling between the dot and the reservoirs, spin flip, etc.) breaking the symmetry, after certain transformation of $\mathcal H$. 

 
At ${E_Z=0}$, the ground state of $\mathcal H$ is the Kondo singlet. It has  entanglement between the impurity spin-1/2 states and reservoir electron states with total spin 1/2. Choosing $z$ axis, the singlet is written as $|\Psi ({E_Z=0}) \rangle = ( | {\uparrow} \rangle | \phi_{-1/2} ({E_Z=0}) \rangle - | {\downarrow} \rangle |\phi_{1/2}  ({E_Z=0}) \rangle )/\sqrt{2}$. $|{\uparrow} (\downarrow) \rangle$ is the impurity state of $S_z = {\uparrow}$ (${\downarrow}$) and  $|\phi_{\mp 1/2}  ({E_Z=0}) \rangle$ is a reservoir state of spin-$z$ quantum number $m_z = \mp 1/2$.
%
When the SU(2) symmetry is broken by $E_Z \ll T_\textrm{K}$ (Boltzmann constant $k_\textrm{B} \equiv 1$), we find~\cite{SUPP}, using bosonization, that the ground state deviates from the Kondo singlet,  
\begin{equation}
 |\Psi (E_Z) \rangle = \alpha_+ (E_Z) | {\uparrow} \rangle |\phi_{-\frac{1}{2}} (E_Z) \rangle - \alpha_-  (E_Z) | {\downarrow}\rangle | \phi_{\frac{1}{2}} (E_Z) \rangle,  \label{GState}
\end{equation}
where  terms of $O(\frac{E_Z^2}{T_\textrm{K}^2})$ are ignored.
$\alpha_\pm = \frac{1}{\sqrt{2}}( 1 \pm \frac{E_Z}{\pi T_\textrm{K}}) + O(\frac{E_Z^2}{T_\textrm{K}^2})$,
$\langle \phi_{1/2} (E_Z) |\phi_{-1/2} (E_Z) \rangle = 0$, and the Anderson orthogonality implies $\langle \phi_{\pm 1/2} (E_Z) |\phi_{\pm 1/2} (E_Z') \rangle = 0$ for $E_Z \ne E_Z'$.
Equation~\eqref{GState} is universal near the Kondo fixed point.
It is a superposition of states of $\langle S_z \rangle + m_z = 0$ since spins are not flipped by $E_Z S_z$.
$|\alpha_+| \ne |\alpha_-|$ means imperfect screening of the impurity spin $\vec{S}$. 

To quantify the imperfect screening, we study entanglement entropy $\mathcal{E}_E \equiv -\textrm{Tr}[\rho_\textrm{D}\log_2 \rho_\textrm{D}]$ between the spin $\vec{S}$ and the reservoirs, where $\rho_\textrm{D} \equiv \textrm{Tr}_\textrm{res} |\Psi (E_Z) \rangle \langle \Psi (E_Z)| $ is the reduced density matrix of the impurity. We find
$\mathcal{E}_E  = 1 - (|\alpha_+|^2 - |\alpha_-|^2)^2 / (2 \log 2)  + O((|\alpha_+|^2 - |\alpha_-|^2)^3) $,  
\begin{eqnarray}
\mathcal{E}_E (E_Z) = 1 - \frac{2}{\pi^2 \log 2} \left( \frac{E_Z}{T_\textrm{K}} \right)^2 + O( \frac{E_Z^4}{T_\textrm{K}^4}). \label{EE1}
\end{eqnarray}
$\mathcal{E}_E$ is maximal at $E_Z=0$, 
and exhibits universal power-law decay with exponent $2$ for small $E_Z$.

The imperfect screening affects the scattering phase shift $\delta_\sigma$ of reservoir electrons with spin $\sigma$ by the dot and their zero-bias conductance  $G_\sigma = G_0 \sin^2(\delta_\sigma)$ through the dot~\cite{Mora} at zero temperature. $G_0 \equiv \frac{e^2}{h}\frac{4\Gamma_\textrm{L} \Gamma_\textrm{R}}{(\Gamma_\textrm{L}+\Gamma_\textrm{R})^2}$ and $\Gamma_{\textrm{L(R)}}$  is the level broadening of the dot state by electron tunneling to the left (right) reservoir.
According to the Fermi liquid theory, we find $\delta_\sigma = \pi [1 +  \chi_\sigma (|\alpha_+|^2 - |\alpha_-|^2)]/2$, 
where $\chi_\uparrow = 1$ and $\chi_\downarrow = -1$. Then the entanglement entropy is related with the total conductance $G_\textrm{T} (E_Z) \equiv G_\uparrow + G_\downarrow = 2G_0 (  1 - E_Z^2/T_\textrm{K}^2 + O(E_Z^4/T_\textrm{K}^4))$ as
\begin{equation}
\mathcal{E}_E(E_Z) = \mathcal{E}_E ({E_Z=0}) - \frac{G_\textrm{T} ({E_Z=0}) - G_\textrm{T} (E_Z)}{ \pi^2 G_0 \log 2} + O( \frac{E_Z^4}{T_\textrm{K}^4}). \label{EE2}
\end{equation}

The result is interesting as a many-body quantum correlation (the entanglement) is determined from a single-particle observable (the conductance). In general, this can happen when a system is in a pure state of a simple form; e.g., a complementary relation~\cite{Jaeger95} connects a single-particle observable and two-particle interference.
In our case, this is a universal property of the Kondo fixed point, near which the ground state has the simple form in Eq.~\eqref{GState}.
Mathematically, the form leads to Eq.~\eqref{EE2}, since both the entanglement and the conductance are functions of only the parameter of $|\alpha_+|^2 - |\alpha_-|^2$.
Physically, nonzero $|\alpha_+|^2 - |\alpha_-|^2$ implies imperfect screening of the impurity spin, which is quantified by $\mathcal{E}_E$ and causes the reduction of $G_\textrm{T}$ from its maximum value.

Equation~\eqref{EE2} holds also when the potential scattering occurs or at finite temperature $T \ll T_\textrm{K}$.
When the particle-hole symmetry of the dot is weakly broken, 
the resulting potential scattering causes additional shift $\delta_\textrm{p} \ll \pi / 2$ in the scattering phase, $\delta_\sigma = \pi [1 +  \chi_\sigma (|\alpha_+|^2 - |\alpha_-|^2)]/2 + \delta_\textrm{p}$.
Then, the conductance changes as
$G_\sigma / G_0 = \sin^2(\delta_\sigma) = 1 - \delta_p^2  - 2\chi_\sigma \delta_p E_Z / T_\textrm{K} - E_Z^2 / T_\textrm{K}^2 + O(E_Z^3 / T_\textrm{K}^3)$  and $G_\textrm{T}/(2G_0) \approx 1-\delta_p^2-E_Z^2/T_\textrm{K}^2$,
while $|\alpha_+|^2 - |\alpha_-|^2  \approx 2 E_Z / (\pi T_\textrm{K})$ and the  entanglement $\mathcal{E}_E$ does not alter.
Hence, Eq.~\eqref{EE2} works.
On the other hand, at finite temperature, the entanglement can be quantified by the entanglement of formation $\mathcal{E}_F$~\cite{ssblee}, a mixed-state generalization of the entanglement entropy. The entanglement follows~\cite{ssblee} $\mathcal{E}_F (E_Z,T) = \mathcal{E}_F (E_Z,T=0) - c_T T^2/T^2_\textrm{K}$ ($c_T >0$ is a constant). The conductance becomes $\mathcal{G}_\textrm{T} (E_Z,T) = \mathcal{G}_\textrm{T} (E_Z,T=0) - 2G_0 (\pi T / T_\textrm{K})^2$. $\mathcal{E}_F (E_Z,T)$ and $\mathcal{G}_\textrm{T} (E_Z,T)$ obey Eq.~\eqref{EE2}.
 
These show that the conductance is useful for detecting the entanglement near the fixed point,
implying that some reported experimental data on quantum dot Kondo effects in fact have information of the entanglement.

Even when the impurity has charge fluctuations, 
the entanglement between the impurity \textit{spin} and the reservoirs satisfies Eq.~\eqref{EE2}.
When the charging energy of the dot is finite (not much larger than $\Gamma_{\textrm{L},\textrm{R}}$), 
the ground state has a charge fluctuation part, e.g., formed by doubly occupied and empty states for the Anderson impurity
or by $|n_\textrm{A}=0, n_\textrm{B}= 0 \rangle$ and $|1,1 \rangle$ for the double dot studied below, 
in addition to the spin part in Eq.~\eqref{GState}.
In this case, Eq.~\eqref{EE1} describes the entanglement between the impurity and the reservoirs in the state obtained by projecting out the charge part~\cite{JShim}.
Meanwhile, the charge part is irrelevant to the dynamics at the Kondo fixed point, including the conductance.
Hence, Eq.~\eqref{EE2} still holds.
%

 


{\it Entanglement in orbital Kondo effects.---} Equation~\eqref{EE2} is useful for detecting the spatial distribution of $\mathcal{E}_E$, an essential information of Kondo cloud.
For the purpose, we propose to use a double dot in Fig.~\ref{setup_AB}. 

In the orbital Kondo regime, the double dot has two degenerate ground states $| {\Rightarrow} \rangle \equiv |n_\textrm{A}=1, n_\textrm{B}= 0 \rangle$ and $| {\Leftarrow} \rangle \equiv |n_\textrm{A}=0, n_\textrm{B}= 1 \rangle$ that act as impurity pseudospin states~\cite{Amasha}.  
The Hamiltonian is $\mathcal H_\textrm{OK} = \mathcal H_\textrm{d}  + \sum_{\eta = \textrm{L}, \textrm{R}; \lambda =\textrm{A}, \textrm{B}} (\mathcal H_\textrm{res}^{\eta \lambda} + \mathcal H^{\eta \lambda}_\textrm{tun})$. 
Here $\mathcal H_\textrm{d} = \epsilon_\textrm{A} n_\textrm{A} + \epsilon_\textrm{B} n_\textrm{B} + U n_\textrm{A} n_\textrm{B}$ describes the double dot.  Each dot $\lambda$ ($= \textrm{A}, \textrm{B}$) is simplified to have a single orbital $d_\lambda^\dagger$ with energy $\epsilon_\lambda < 0$ and electron occupation number $n_\lambda = d_\lambda^\dagger d_\lambda$. $U$ is the interdot Coulomb energy.
$\mathcal H_\textrm{res}^{\eta \lambda} = - t \sum_{ j=1}^\infty c_{j, \eta\lambda}^\dagger  c_{j+1, \eta\lambda}+ \textrm{h.c.}$ describes reservoir $\eta\lambda$ ($\eta =$ L, R). $c^\dagger_{j,\eta\lambda}$ creates an electron in the site $j$ of the reservoir, $t$ is the hopping energy, and $\textrm{h.c.}$ means hermitian conjugate.
$\mathcal H^{\eta \lambda}_\textrm{tun} = - t_{\eta \lambda} d_\lambda^\dagger c_{1,\eta \lambda } + \textrm{h.c.}$ describes electron tunneling between dot $\lambda$ and its own reservoir $\eta \lambda$ with strength $t_{\eta \lambda}$, leading to dot-level broadening $\Gamma_{\eta \lambda}$.
We consider the orbital Kondo regime of $U \gg \Gamma_{\eta \lambda}$.
We focus on the symmetric case of $\epsilon_\textrm{A} = \epsilon_\textrm{B} = -U/2$ and $t_{\eta \textrm{A}} = t_{\eta \textrm{B}}$ where the orbital Kondo effect maximally occurs.
We ignore electron spin, considering a magnetic field destroying spin Kondo effects.

In addition, we consider electron tunneling between reservoirs $\eta\textrm{A}$ and $\eta\textrm{B}$, with strength $t_\textrm{sf}$, in the region outside distance $l$ from the dot. Its Hamiltonian is
\begin{eqnarray}
\mathcal H_\textrm{sf} &=& \sum_{\eta = \textrm{L}, \textrm{R}} \mathcal H_\textrm{sf}^\eta = - \sum_{\eta = \textrm{L}, \textrm{R}} \sum_{j=l}^\infty  t_\textrm{sf} \, c_{j,\eta \textrm{A}}^\dagger c_{j,\eta \textrm{B}} + \textrm{h.c.}  \label{Hsb}
\end{eqnarray}
It breaks the SU(2) pseudospin symmetry of $\mathcal H_\textrm{OK}$ spatially nonuniformly in the reservoirs.
 Note that our main results do not alter when the tunneling $t_\textrm{sf}$ turns on only between LA and LB or between RA and RB as in Fig.~\ref{setup_AB}. 
 
\begin{figure}[tb]
\includegraphics[width=\columnwidth]{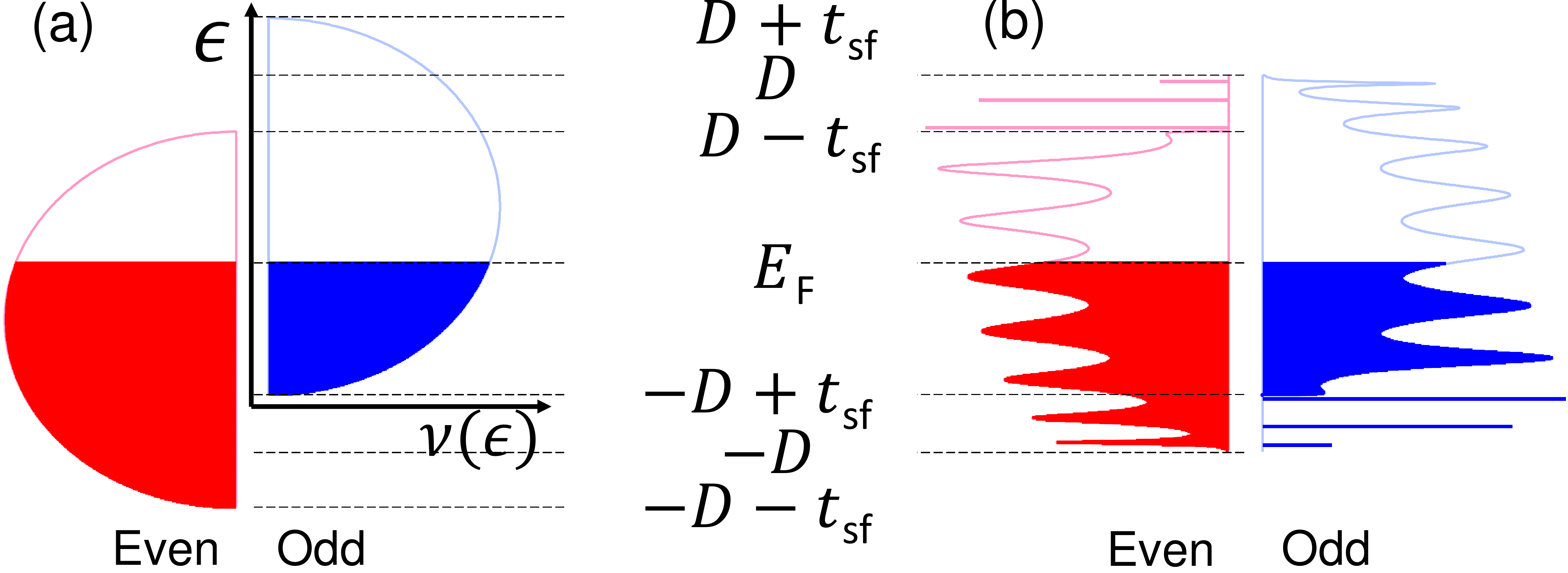}
\caption{Energy dependence of local densities of states (LDOS) $\nu (\epsilon)$ at reservoir sites adjacent to the double dot. The LDOS of even reservoir states $c^\dagger_{k, \eta \textrm{E}}$ (odd $c^\dagger_{k, \eta \textrm{O}}$) is marked by E (O).
Occupied (unoccupied) states are shown by filled (empty) regions.
(a) The case of $l = 1$, where the inter-reservoir tunneling occurs uniformly over the whole region. (b) The $l > 1$ case, where the tunneling occurs outside the distance $l$. The LDOS has resonance or localized-state peaks. 
} \label{LDOS}
\end{figure}

The ground state of the total Hamiltonian $H_\textrm{tot} = H_\textrm{OK} + \mathcal{H}_\textrm{sf}$ has the form in Eq.~\eqref{GState}. To see this, we first consider the ${l = 1}$ case where the inter-reservoir tunneling occurs uniformly over the whole region. We use even-odd superpositions of A and B, the dot states of $|{\Uparrow} \rangle = (|{\Rightarrow} \rangle + |{\Leftarrow} \rangle) / \sqrt{2}$,  $|{\Downarrow} \rangle = (|{\Rightarrow} \rangle - |{\Leftarrow} \rangle) / \sqrt{2}$, and the reservoir operators of $c^\dagger_{j, \eta \textrm{E}} = (c^\dagger_{j, \eta \textrm{A}} + c^\dagger_{j, \eta \textrm{B}})/\sqrt{2}$ and $c^\dagger_{j, \eta \textrm{O}} = (c^\dagger_{j, \eta \textrm{A}} - c^\dagger_{j, \eta \textrm{B}})/\sqrt{2}$. Then, our setup is viewed as an impurity pseudospin (its $S_z$ states are $\Uparrow$ and $\Downarrow$) coupled with a ferromagnetic reservoir [see Fig.~\ref{LDOS}(a)] where the even (odd) modes $c^\dagger_{j, \eta \textrm{E(O)}}$ support majority (minority) pseudospin states, as the reservoir Hamiltonian becomes 
$\mathcal H_\textrm{res}^{\eta \textrm{A}} + \mathcal H_\textrm{res}^{\eta \textrm{B}} + \mathcal{H}_\textrm{sf}^\eta = \sum_{k} (\epsilon_k - t_\textrm{sf})  c_{k, \eta \textrm{E}}^\dagger c_{k, \eta \textrm{E}} + (\epsilon_k + t_\textrm{sf} ) c_{k, \eta \textrm{O}}^\dagger c_{k, \eta \textrm{O}}$ after Fourier transforming  $c_{j, \eta \textrm{E}}$ ($c_{j, \eta \textrm{O}}$) into $c_{k, \eta \textrm{E}}$ ($c_{k,\eta \textrm{O}}$).
$\epsilon_k= - 2 t \cos k a$ and $a$ is the lattice spacing. When $t_\textrm{sf} \ll T_\textrm{K}$, the ground state is 
 \begin{equation}
|\Psi_\textrm{OK} (t_\textrm{sf}) \rangle= \alpha_+  |\Uparrow \rangle|\varphi^\textrm{EO}_{m_z=-1/2}  \rangle - \alpha_-  |\Downarrow \rangle |\varphi^\textrm{EO}_{m_z= 1/2}  \rangle.  \label{GStateEO}
\end{equation}
$\alpha_\pm = \frac{1}{\sqrt{2}}( 1 \pm \frac{2t_\textrm{sf}}{\pi T_\textrm{K}}) + O(\frac{t_\textrm{sf}^2}{T_\textrm{K}^2})$
and $|\varphi^\textrm{EO}_{m_z}  (t_\textrm{sf})  \rangle$ is a reservoir state (written by $c_{k, \eta \textrm{E}}$ and $c_{k, \eta \textrm{O}}$) with pseudospin-z quantum number  $m_z$.
It has the same form with Eq.~\eqref{GState} except replacement $E_Z \to 2 t_\textrm{sf}$~\cite{SUPP}, as
our setup is also viewed as an impurity spin with Zeeman splitting $2 t_\textrm{sf}$ coupled with a non-magnetic reservoir, described by Eq.~\eqref{Hamiltonian}. 
Accordingly, the entanglement between the impurity pseudospin and the reservoirs satisfies  Eq.~\eqref{EE1}, 
\begin{equation}
\mathcal E_E (t_\textrm{sf}) = 1 - \frac{2}{\pi^2 \log 2} \left( \frac{2 t_\textrm{sf}}{T_\textrm{K}} \right)^2 + O(\frac{t_\textrm{sf}^4}{T_\textrm{K}^4}). \label{EE_EO}
\end{equation}

{\it Spatial distribution of the entanglement.---} We move to the $l > 1$ case. The inter-reservoir tunneling now occurs outside the distance $l$. We will study the $l$ dependence of the entanglement $\mathcal{E}_E$ and the conductance $\mathcal{G}_\textrm{T}$ through the double dot from the left reservoirs $\textrm{L} \lambda$'s to the right $\textrm{R} \lambda$'s, and show that $\mathcal{E}_E(l)$ and $\mathcal{G}_\textrm{T}(l)$ satisfy Eq.~\eqref{EE2}.

For the purpose, we use the even-odd bases and study the local densities of states  (LDOS) $\nu_\textrm{E(O)}$ of even (odd) reservoir states $c^\dagger_{j, \eta \textrm{E(O)}}$ at the sites adjacent to the double dot. 
In the bases, the inter-reservoir tunneling makes the energy band of the even (odd) states outside the distance $l$ shift downward (upward) by $t_\textrm{sf}$. Hence the even (odd) states in energy window $[D - t_\textrm{sf}, D]$ ($[-D, -D + t_\textrm{sf}]$) are localized states within $l$ and provide discrete LDOS peaks, where $2D=4t$ is the band width. Those in the other window are resonance states, resulting in continuous LDOS with broadened peaks [see Fig.~\ref{LDOS}(b)].  
The contribution of the resonance states to the LDOS is found as (that of the localized states is not shown)
\begin{eqnarray}
\nu_\textrm{E/O}(\epsilon) = \frac{1}{\pi t} \frac{\sin(qa)}{1 \pm \frac{t_\textrm{sf}}{t} \frac{\sin(k(l-1)a)  \sin(kla)}{\sin^2(ka)}}, \label{eq:ldos}
\end{eqnarray}
where the energy $\epsilon$, wave vector $k$ inside $l$, and wave vector $q$ outside $l$ satisfy $\epsilon = -2t\cos(ka) = -2t\cos(qa) \mp t_\textrm{sf}$ and the upper (lower) sign is for the even (odd) states.  




The difference of the LDOS between the even reservoir states and the odd ones weakens the orbital Kondo effect.
It leads to the difference $\Delta n_\textrm{res} \equiv n_\textrm{res,E}-n_\textrm{res,O}$ of their electron occupation number~\cite{SUPP},
\begin{eqnarray} 
\Delta n_\textrm{res}= (-1)^{l+1} \frac{2 t_\textrm{sf}}{\pi D}\frac{1}{l}  + O(\frac{t^2_\textrm{sf}}{l^2}). \label{deltaNres}
\end{eqnarray}
The occupation number of the even (odd) reservoir states is $n_\textrm{res,E(O)}\equiv \int^{E_F}_{-D} \nu_\textrm{E(O)}(\epsilon) d\epsilon$, and $E_\textrm{F} = 0$ is the Fermi level.
According to the Fermi liquid theory~\cite{Mora,Affleck_CFT1},
this induces the difference $\Delta n_\textrm{dot} \equiv n_\textrm{dot,E} - n_\textrm{dot,O}$ of the occupation between the double-dot states $|{\Uparrow} \rangle$ and $| {\Downarrow} \rangle$ as $\Delta n_\textrm{dot} = 4 c \Delta n_\textrm{res} D / (\pi T_\textrm{K})$, where the occupation of $|{\Uparrow} \rangle$ ($| {\Downarrow} \rangle$) is $n_\textrm{dot,E(O)}\equiv \int_{-D}^{E_F} \mathcal A_\textrm{E(O)}(\epsilon) d\epsilon$ and $\mathcal A_\textrm{E(O)}$ is the impurity spectral function for $\Uparrow$ ($\Downarrow$). $c$ is a constant of $O(1)$, and $c=1$ when the LDOS is energy independent. On the other hand, $\Delta n_\textrm{dot}=|\alpha_+|^2 - |\alpha_-|^2$, because of the state form in Eq.~\eqref{GStateEO}. Following the steps discussed around Eqs.~\eqref{EE1} and \eqref{EE2}, we derive the entanglement entropy between the impurity pseudospin and the reservoirs, and the conductance through the double dot as 
\begin{eqnarray}
\mathcal E_E (t_\textrm{sf},l) & = & 1 - \frac{2 c^2}{\pi^2\log 2} \left( \frac{4 t_\textrm{sf}}{\pi D} \right)^2 \left( \frac{\xi_\textrm{K}}{la} \right)^2  + O( \left( \frac{t_\textrm{sf} \xi_\textrm{K} }{D l a} \right)^4 ), \nonumber\\
\frac{\mathcal G_\textrm{T} (t_\textrm{sf},l)}{2 \mathcal{G}_0}  & = & 1 -   c^2 \left( \frac{4 t_\textrm{sf}}{\pi D} \right)^2 \left( \frac{\xi_\textrm{K}}{la} \right)^2 + O( \left( \frac{t_\textrm{sf} \xi_\textrm{K} }{D l a} \right)^4 ). \label{ent and cond form}
\end{eqnarray}
$\xi_\textrm{K}\equiv \hbar v_F / T_\textrm{K}$ is the Kondo cloud length,  $v_F = 2 t a = Da$ is the Fermi velocity, and $\mathcal{G}_0  \equiv \frac{e^2}{h}\frac{4\Gamma_\textrm{LA} \Gamma_\textrm{RA}}{(\Gamma_\textrm{LA}+\Gamma_\textrm{RA})^2}= \frac{e^2}{h}\frac{4\Gamma_\textrm{LB} \Gamma_\textrm{RB}}{(\Gamma_\textrm{LB}+\Gamma_\textrm{RB})^2}$.

\begin{figure}[tb]
\includegraphics[width=\columnwidth]{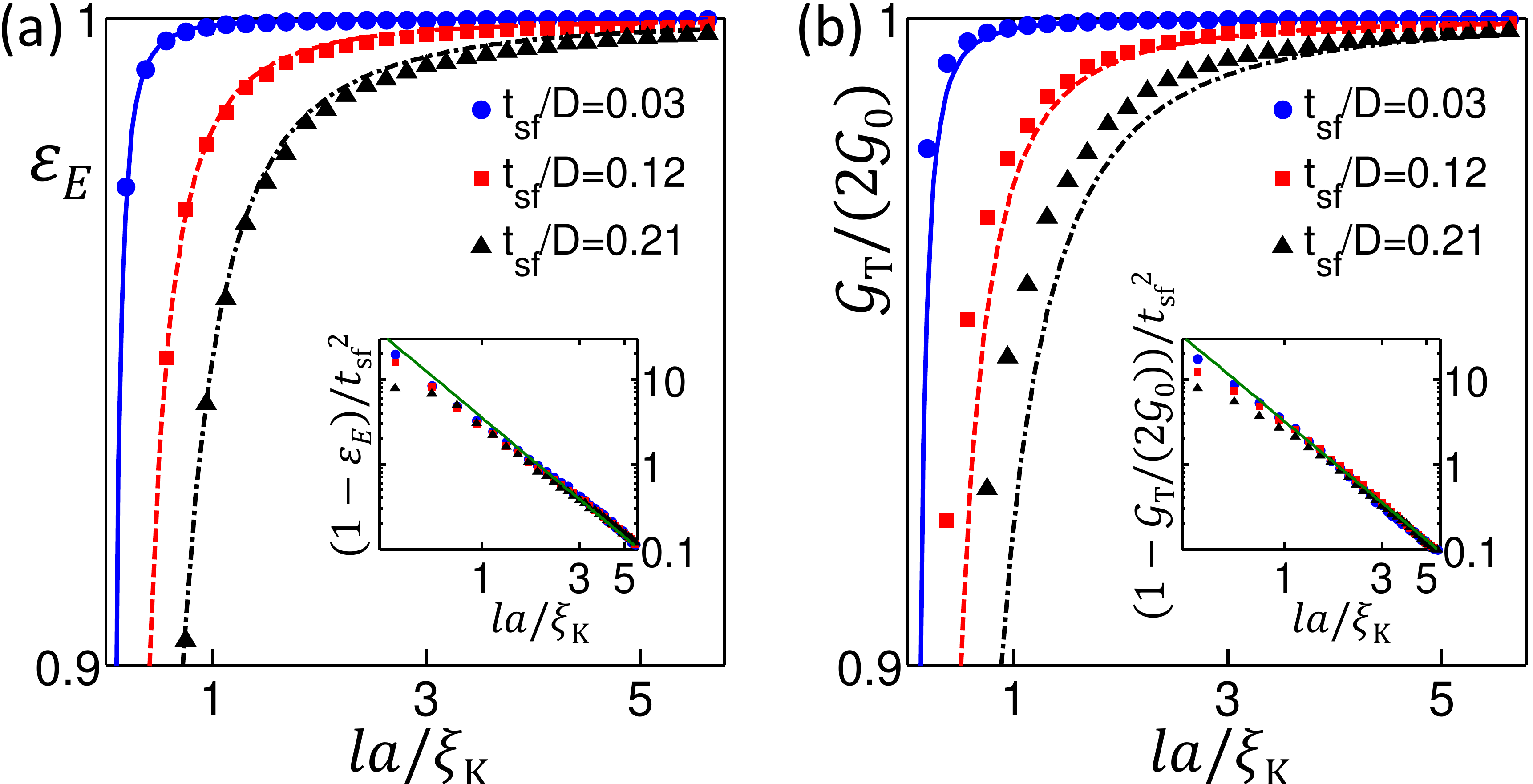}
\caption{NRG results (symbols) of (a) the entanglement entropy $\mathcal{E}_E$ and (b) the conductance $\mathcal{G}_\textrm{T}$ through the double dot as a function of $la / \xi_\textrm{K}$ for $t_\textrm{sf} / D = 0.03, 0.12, 0.21$. The results agree with the curves representing $\mathcal E_E=1 - \frac{2}{\pi^2 \log 2} \frac{(4t_\textrm{sf})^2}{(\pi D)^2} \frac{\xi_\textrm{K}^2}{l^2 a^2}$ and  $\frac{\mathcal{G}_\textrm{T}}{2 \mathcal{G}_0} = 1 - \frac{(4t_\textrm{sf})^2}{(\pi D)^2} \frac{\xi_\textrm{K}^2}{l^2 a^2}$ for small $\xi_\textrm{K}  t_\textrm{sf} / (la D)$. 
The insets are the log-log plots of  the dependence of $(1 - \mathcal{E}_E) / t_\textrm{sf}^2$ and $(1 - \mathcal{G}_\textrm{T} / (2 \mathcal{G}_0))/ t_\textrm{sf}^2$ on $la / \xi_\textrm{K}$. The results follow the linear curve (green lines) of slope $-2$, meaning that $(1 - \mathcal{E}_E) / t_\textrm{sf}^2$ and $(1 - \mathcal{G}_\textrm{T} / (2 \mathcal{G}_0))/ t_\textrm{sf}^2$ are proportional to $(la / \xi_\textrm{K})^{-2}$.
} \label{NRG}
\end{figure}

To confirm Eq.~\eqref{ent and cond form}, we perform NRG calculations~\cite{SUPP} for various values of $t_\textrm{sf}$, choosing $U/D = 3.6$ and $t_{\eta \lambda}/D = 0.34$.
The result in Fig.~\ref{NRG} shows $1 - \mathcal{E}_E \propto (t_\textrm{sf} \xi_\textrm{K} / l)^2$ and $1 - \mathcal{G}_\textrm{T} / (2 \mathcal{G}_0) \propto  (t_\textrm{sf} \xi_\textrm{K} / l)^2$ in good agreement~\cite{note_error} with Eq.~\eqref{ent and cond form} for sufficiently small $t_\textrm{sf}$ and large $l$.

There are interesting implications of Eq.~\eqref{ent and cond form}. First, the entanglement shows the power-law decay with exponent $-2$ as a function of the distance $l$. It means that the Kondo cloud has a long tail of algebraic decay.
Note that the exponent is different from the exponent $-1$ of the distance dependence of the entanglement obtained~\cite{ssblee} by tracing out the reservoir outside the distance (instead of pseudospin flip by the inter-reservoir tunneling in this study).
Second, the entanglement and the conductance in Eq.~\eqref{ent and cond form} satisfy Eq.~\eqref{EE2}. Hence the power-law decay of the entanglement can be detected by measuring the $l$ dependence of the conductance.
Third, when the inter-reservoir tunneling is large as $t_\textrm{sf} \simeq D/2$, $\mathcal{G}_\textrm{T} / (2 \mathcal{G}_0) \simeq 0.9$ at $l a = 2 \xi_\textrm{K}$. By using this, one can estimate the Kondo cloud length in experiments.
Forth, $\mathcal{E}_E$ in Eq.~\eqref{ent and cond form} is applicable to the core region $l a \lesssim \xi_\textrm{K}$ of the Kondo cloud, provided $t_\textrm{sf} < D /2$. The suppression of the cloud due to the $l$-dependent SU(2) symmetry breaking follows the same power law of exponent $-2$, reflecting  the Fermi liquid, both in the core and the tail of the cloud.

{\it Discussion.---} We have found that the entanglement between a Kondo impurity spin and electron reservoirs can be determined by electron conductance through a quantum dot in the single-channel Kondo regime. The power law in Eq.~\eqref{EE1} is valid for perturbations breaking the SU(2) symmetry. Similar behavior is expected for other observables (such as spin susceptibility, heat capacitance, and local density of states~\cite{Affleck_attempt1,Bergmann,Busser,Ribeiro}), for other entanglement measures (such as R\'{e}nyi entropies and entanglement negativity), and for other setups. Equation~\eqref{EE2} is a simple relation but has been unnoticed before. Although it is applicable only to the regime near the Kondo fixed point, Eq.~\eqref{EE2} will be useful for detecting many-body entanglement in various systems that can be mapped onto the Kondo model in Eq.~\eqref{Hamiltonian}. 

Our strategy for detecting a Kondo cloud based on Eqs.~\eqref{EE2} and \eqref{ent and cond form}
is within experimental reach.
For example, the orbital Kondo effect was measured in a double dot~\cite{Amasha} that has two reservoirs separated by a barrier formed by an electrical gate. The reservoirs correspond to those ($\eta \textrm{A}$ and $\eta \textrm{B}$) of our setup.
By replacing the gate by key-board type gates, the $l$-dependent inter-reservoir tunneling can be tuned.
In this case, it is required that the length scale $\delta l$ over which $t_\textrm{sf}$ changes from 0 to a constant value is shorter than the Fermi wave length $\lambda_\textrm{F}$, not to wash out the resonances near $E_\textrm{F}$.
This can be achieved in semiconductor two-dimensional electron systems of long $\lambda_\textrm{F}$ (as in Ref.~\cite{Amasha}) or when the band bottom of a sub-band channel of the reservoirs lies slightly below $E_\textrm{F}$. 
Our strategy works at finite temperature $T \ll T_\textrm{K}$ or when potential scattering exists, as discussed before.
It also works when the symmetry between A and B is broken, provided that the resulting pseudospin Zeeman splitting is smaller than $T_\textrm{K}$. 
Our strategy is distinct from the existing proposals for detecting Kondo cloud~\cite{Affleck_cloud1,Affleck_attempt2,Cornaglia,jinhong,Simon03,Hand}, as it pursues to detect the non-classical nature (entanglement) of the cloud and it is not to extract the Kondo cloud length from the temperature dependence of an observable based on $\xi_\textrm{K} = \hbar v_\textrm{F} / T_\textrm{K}$.
 
It is remarkable that a many-body entanglement in electron systems can be detected by a single-particle observable.
It will be valuable to generalize our study to other quantum impurity problems. Note that the entanglement studied in this work is different from the impurity entanglement entropy~\cite{Sorensen2,Eriksson}. It will be interesting to find a relation between the impurity entropy and certain observables.
 

We thank Yunchul Chung, David Goldhaber-Gordon, Lucas Peeters, and especially Eran Sela for valuable discussions, and the support by Korea NRF (Grant Nos. 2015R1A2A1A15051869 and 2016R1A5A1008184).


\begin{thebibliography}{}
\bibitem{note_error} In Fig.~\ref{NRG}, the analytic expressions of Eq.~\eqref{ent and cond form} with $c=1$ become to deviate from the NRG results for smaller $la / \xi_\textrm{K} \lesssim 1$ and larger $t_\textrm{sf} / D \gtrsim 0.1$. There are two reasons: (i) The fourth-order terms $[4t_\textrm{sf} \xi_\textrm{K} / (\pi D la)]^4$ in Eq.~\eqref{ent and cond form} become non-negligible, as indicated in the insets of Fig.~\ref{NRG}. The conductance has a larger coefficient (by factor $2.5$) in the fourth-order term than the entanglement entropy, resulting in larger deviation.
(ii) The conductance is obtained from the height of the spectral function at $E_\textrm{F}$.
When the LDOS is more energy-dependent, the structure (e.g., slope, curvature) of the LDOS more affects the height.
In this case, the NRG result of the height can become more sensitive to numerical artifacts.
\end{thebibliography}
\end{document}